\documentclass{aa}  
\usepackage{graphicx,color}
\usepackage{txfonts}
\usepackage{amsmath}
\usepackage{xcolor}
\usepackage{comment}
\usepackage{natbib,twoopt}
\usepackage[hyphenbreaks]{breakurl}
\usepackage[breaklinks]{hyperref}  
\DeclareMathAlphabet\mathzapf       {T1}{pzc} {mb} {it}
\bibpunct{(}{)}{;}{a}{}{,}             
\definecolor{cobalt}{rgb}{0.06, 0.2, 0.65}
\hypersetup{
  colorlinks,
  citecolor=cobalt,
  linkcolor=cobalt,
  urlcolor=cobalt,
}
\makeatletter
  \newcommandtwoopt{\citeads}[3][][]{\href{http://adsabs.harvard.edu/abs/#3}%
    {\def\hyper@linkstart##1##2{}%
     \let\hyper@linkend\@empty\citealp[#1][#2]{#3}}}
  \newcommandtwoopt{\citepads}[3][][]{\href{http://adsabs.harvard.edu/abs/#3}%
    {\def\hyper@linkstart##1##2{}%
     \let\hyper@linkend\@empty\citep[#1][#2]{#3}}}
  \newcommandtwoopt{\citetads}[3][][]{\href{http://adsabs.harvard.edu/abs/#3}%
    {\def\hyper@linkstart##1##2{}%
     \let\hyper@linkend\@empty\citet[#1][#2]{#3}}}
  \newcommandtwoopt{\citeyearads}[3][][]%
    {\href{http://adsabs.harvard.edu/abs/#3}
    {\def\hyper@linkstart##1##2{}%
     \let\hyper@linkend\@empty\citeyear[#1][#2]{#3}}}
\makeatother
\newcommand{\be}{\begin{equation}}
\newcommand{\en}{\end{equation}}

\def\ltsima{$\; \buildrel < \over \sim \;$}
\def\lsim{\lower.5ex\hbox{\ltsima}}
\def\gtsima{$\; \buildrel $\geq$ \over \sim \;$}
\def\gsim{\lower.5ex\hbox{\gtsima}}

\def\heii {He~$\textsc{ii}$}
\def\hei {He~$\textsc{i}$}
\def \ha {H$\alpha$}
\def \hb {H$\beta$}

\begin{document} 
   \title{Accretion disc winds imprint distinct signatures in the optical variability spectrum of black hole transients}
%\title{Accretion disc winds imprint distinct signatures in the optical rms spectrum of black hole tfransients}

\titlerunning{The imprint of winds in the rms optical spectrum}
\authorrunning{Vincentelli \& Muñoz-Darias}

   \author{Federico~M.~Vincentelli
          \inst{\ref{i1},\ref{i2},\ref{i3},\ref{i4}}
          \and
          Teo~Muñoz-Darias\inst{\ref{i1},\ref{i2}}
          }

\institute{
Instituto de Astrofísica de Canarias, E-38205 La Laguna, Tenerife, Spain \label{i1}
 \and
Departamento de Astrofísica, Universidad de La Laguna, E-38206 La Laguna, Tenerife, Spain \label{i2}
\and
School of Physics \& Astronomy, University of Southampton, Southampton
SO17 1BJ, UK\label{i3}
\and
INAF-Osservatorio Astronomico di Roma, via Frascati 33, I-00078 Monteporzio Catone (RM), Italy\label{i4}}

   %\date{Received ...; accepted ...}
\date{}
% \abstract{}{}{}{}{} 
% 5 {} token are mandatory

\abstract{Quantifying the variability, measured as the root mean square (rms), of accreting systems as a function of energy is a powerful tool for constraining the physical properties of these objects. Here, we present the first application of this method to optical spectra of low-mass X-ray binaries. We use high-time-resolution data of the black hole transient V404 Cygni, obtained with the \textit{Gran Telescopio Canarias} during its 2015 outburst. During this event, conspicuous wind-related features, such as P-Cygni profiles, were detected in the flux spectra.

We find that rms spectra are rich in spectral features, although they are typically morphologically different from their flux counterparts. Specifically, we typically observe absorption components in correspondence to the presence of emission lines in the flux spectra. Similarly, when analysing segments with significant variability in the optical flux, P-Cygni line profiles appear inverted in the rms spectra (i.e., enhanced variability in the blue-shifted region, accompanied by a decrease in that associated with the red component).
We discuss the possible origin of these features, which resemble those found in other objects, such as active galactic nuclei. Finally, we highlight the potential of this technique for future searches for wind-type outflows in accreting compact objects.}


   \maketitle
%
%-------------------------------------------------------------------

\section{Introduction}

Low-mass X-ray binaries (LMXBs) host a compact object (either a black hole or a neutron star) that is accreting mass from a low-mass star via an accretion disc. A large fraction of them are transient systems, spending most of their time in a quiescent state. This is interrupted by short periods of intense activity (weeks to years), known as outbursts, during which they can increase their luminosity by more than a million times reaching up to $\sim 10^{37-39}$ erg s$^{-1}$. LMXBs are truly multi-wavelength sources, with a very broad and highly variable spectral energy distribution, ranging  from hard X-rays to radio \citep[see e.g.][]{corbel2002,gandhi2011,migliari2010,russell_t2013,diaztrigo2017-1728,diaz-trigo2018,tetarenko2021_a}.  Such a complex phenomenology is the result of different emitting components changing their contribution throughout the outburst. 

In X-rays, the spectrum can be dominated either by a thermal component, arising in a geometrically thin, optically thick accretion disk (peaking in the soft X-rays), or  by a hard non-thermal  spectrum associated with an optically thin, geometrically thick Comptonising medium \citep{zdariskigierlinski}. The relative contribution of these components changes significantly during the outburst, defining different spectral states (soft and hard). Interestingly, the different X-ray spectral states are accompanied by a set of well-defined multi-wavelength and timing properties (e.g. \citealt{McClintock2006, belloni2010,Fender2016}) . In the hard state, a strong stochastic aperiodic noise on second and sub-second timescales is observed in X-rays, in addition to synchrotron emission from a jet that dominates the radio and the infrared, and even makes a contribution to the to optical regime. In the soft state, the X-ray variability is {strongly reduced \citep[see e.g.][]{Munoz-Darias2011,Munoz-Darias2014}}, and the radio jet is, at least, significantly quenched (e.g. \citealt{fender2004,Russell2011}). In addition to jets, wind-type outflows are observed primarily through blue-shifted absorption lines detected across various electromagnetic bands. These outflows are seen in X-rays (e.g., \citealt{neilsen2009, ponti2012, Parra2024}), optical (e.g., \citealt{Munoz-Darias2019, PanizoEspinar2022}), and more recently in the near-infrared (e.g., \citealt{Sanchez-Sierras2023b}) and far-ultraviolet bands (e.g., \citealt{castro-segura_2022,Fijma2023}). A detailed analysis of this complex phenomenology suggests that it can be explained by a clumpy, multi-phase outflow that is consistently present and changes its properties based on several physical factors, such as the intensity and spectral energy distribution of the incident radiation \citep{sanches-sierras2020, castro-segura_2022, Munoz-Darias2022}

\begin{table*}[]

\centering
\caption{GTC data used in our study.}
\begin{tabular}{|c|cc|c|}
\hline
Dataset & \# of spectra &   Duration (min.)  & Lightcurve properties\\
\hline
&&&\\
day 1   & 37 & 52 & \textbf{decay} of more than 1 mag in  1 h  \\
&&&\\
day 2   & 76 & 102 & \textbf{double-peaked flare}: including its rise, peak  and decay \\
&&&\\
day 3   & 76 & 102  & \textbf{slow rise} in the continuum flux\\
&&&\\

day 4   & 76 & 102 & \textbf{slow decay}: brighter than day 1 \\
&&&\\

day 5   & 41 & 54 & \textbf{decay  ending in a plateau}  \\
&&&\\

day 6   & 86 & 117 & \textbf{full flare}: Brighter than day 2, with faster decay  \\
&&&\\

day 7   & 37 & 48 & \textbf{flares}: fastest ($<$ 5 min) but also weakest  \\

&&&\\

day 8   & 37 & 48 & \textbf{bright plateau}  \\

\hline
\end{tabular}
\label{tab:data}
  \tablefoot{Along with the day of the outburst (starting from MJD 57189), we provide the number of spectra, the total duration of the observation, and a description of the lightcurve. Further details on each dataset can be found in \citet{Munoz-Darias2016} and \citet{mata-sanchez2018}. }
\end{table*}

A handful of LMXBs have shown extreme variability during some phases of their outburst, displaying  recurrent flares on timescales of minutes, which in some cases has been associated with a radiation pressure instability occurring close to the Eddington Luminosity  \citep{belloni1997,janiuk2000,neilsen2011,vincentelli2023,wang_jingy_2024}.
One of such systems, the black hole transient V404~Cyg, displayed a luminous outburst in 2015, that showed remarkable observational properties (e.g. \citealt{kimura2016,motta2017b,Fender2023}) including the presence of massive winds observed in the optical (\citealt{Munoz-Darias2016, Munoz-Darias2017, Casares2019}) and X-ray (\citealt{King2015,motta2017b}) domains. In particular, the optical spectroscopy obtained during the outburst has an unprecedented high-cadence ($\sim$1 minute; see e.g. \citealt{mata-sanchez2018}) which enables to study in great detail the variability properties of the conspicuous wind signatures. In this paper, we attempt to quantify the wavelength-dependent optical variability characteristic of extreme accretion phases by analysing {root-mean-squared} (rms) spectra computed from this unique dataset. 

\section{Data} 
\label{sec:data}
In this paper we make use of the spectroscopy taken with the \textit{Gran Telescopio Canarias} during the 2015 outburst of V404 Cyg. The data set has already been presented in \citet{Munoz-Darias2016} and \citet{mata-sanchez2018}, to which we refer the reader for further details on the observations, including the data reduction. The campaign lasted about two weeks, during which more than 500 spectra were taken. It covered the most active phase of the outburst. We focused our attention on the  data of from the first eight days (hereafter day 1 to day 8; see Table \ref{tab:data}). During this period, the source displayed some of the strongest optical/near-infrared wind signatures ever detected in black hole transients (see also \citealt{Munoz-Darias2018, Sanchez-Sierras2023b}). Notably, these observations include P-Cygni line profiles with blue-shifted absorptions reaching up to $\sim$30 percent below the continuum level, and terminal velocities (defined as the blue edge of the absorption component) ranging from $\sim$1500 to 3000 km s$^{-1}$. Each observation contains 37 to 86 spectra, taken with regular sampling over periods of 48 to 117 minutes—a timescale on which the source exhibits conspicuous variability that has been found to correlate with the observational properties of the wind \citep{Munoz-Darias2016}. This, combined with the high signal-to-noise ratio (S/N) of the individual spectra and the accurate flux calibration of the dataset, provides a unique opportunity to compute rms optical spectra using the following methodology.

\begin{figure*}
\centering 

\includegraphics[width=0.7\textwidth]{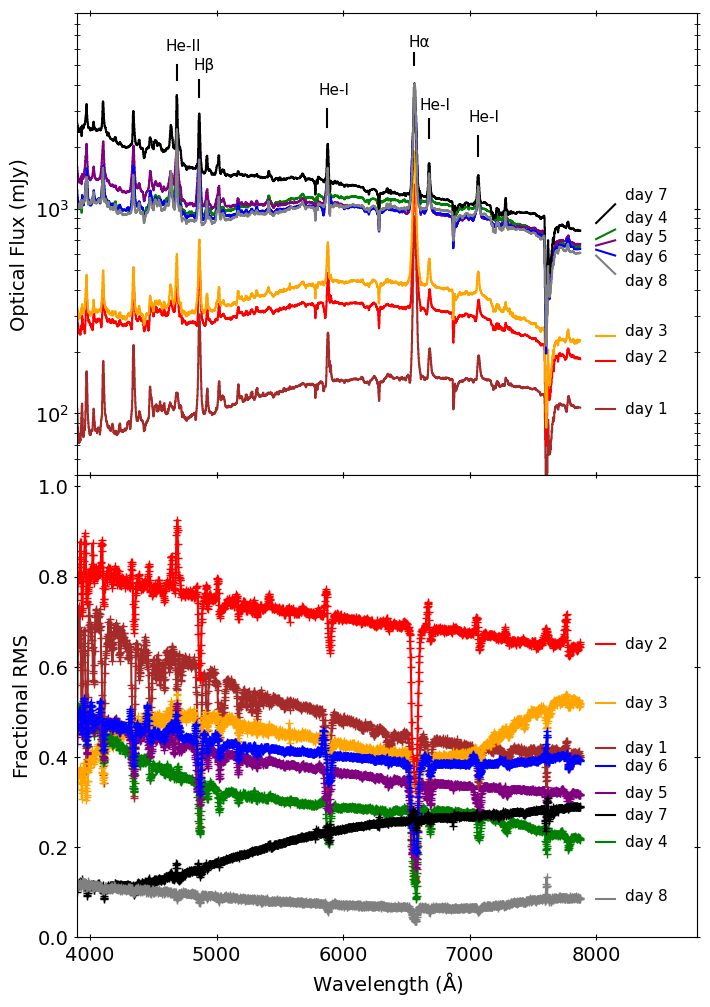}

\caption{Flux-averaged (top) and fractional rms (bottom) spectra for the eight epochs included in this study. In the top panel, we mark the main lines that will be analysed and discussed in this paper: \hei\ lines at $\sim$ 5876, 6678 and 7065 \AA, \heii\--4686, \ha\ and \hb.} 
    \label{fig:all}

\end{figure*}

\begin{figure*}
    \centering
\includegraphics[width=0.45\textwidth]{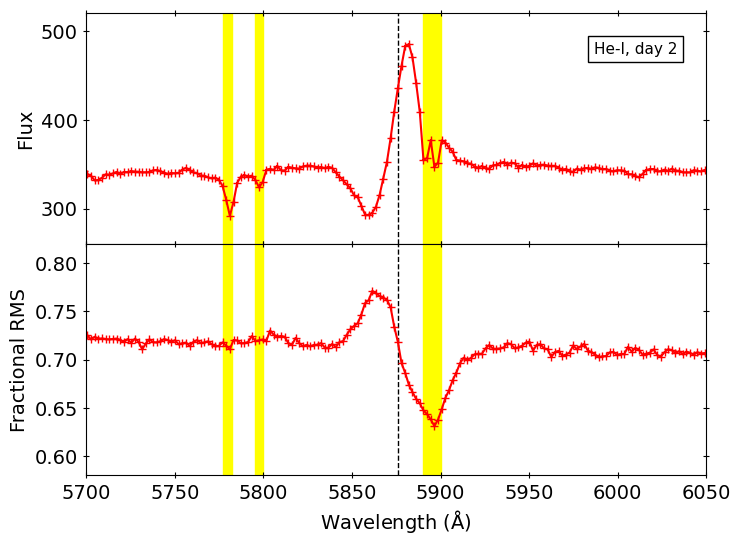}
\includegraphics[width=0.45\textwidth]{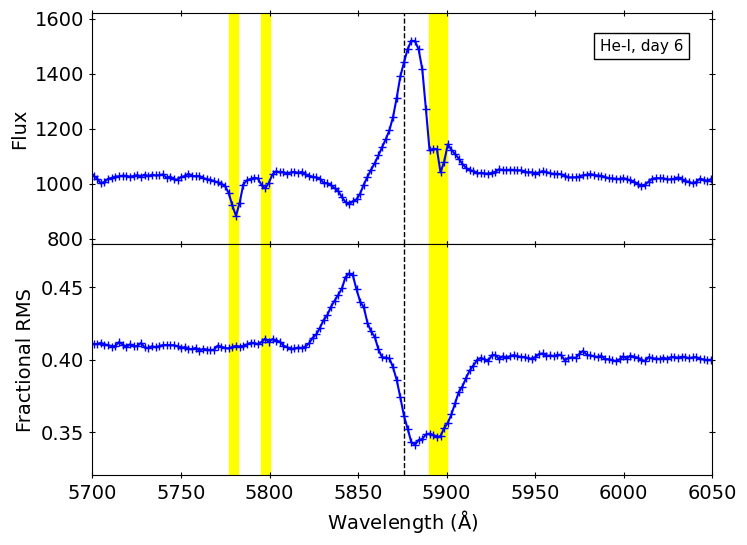}

\includegraphics[width=0.45\textwidth]{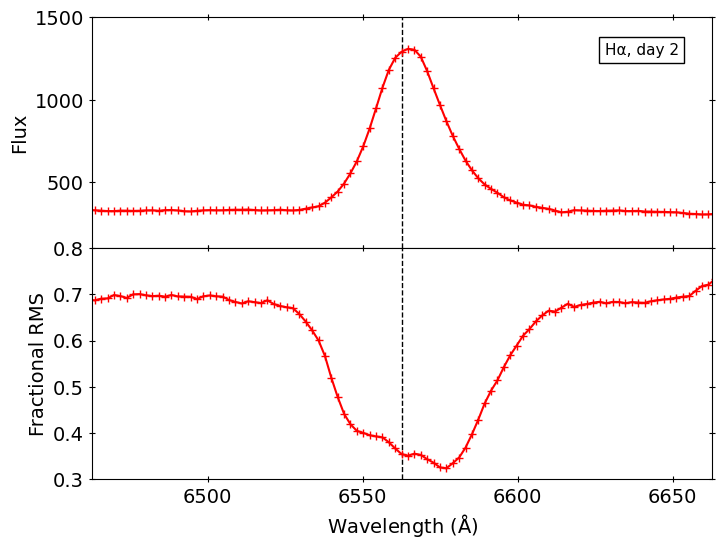}
\includegraphics[width=0.45\textwidth]{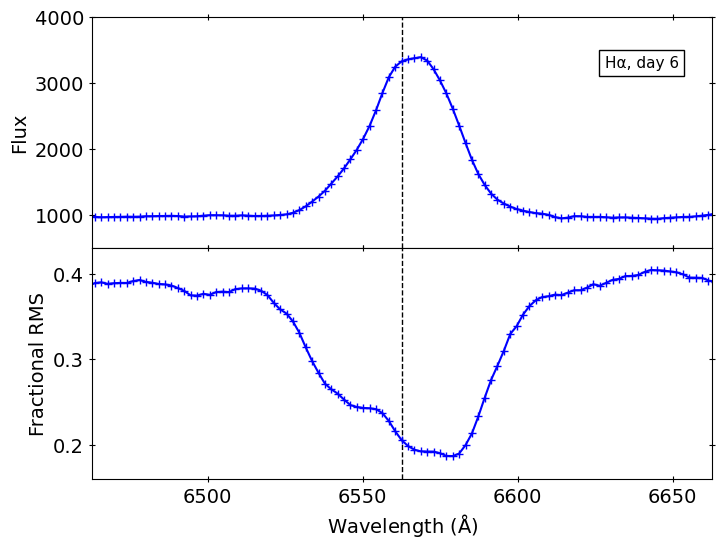}

\caption{Detailed rms and flux spectra from day 2 and day 6. The four panels show the flux (top) and fractional rms (bottom) spectra for day 2 (left) and day 6 (right). The top row presents the behaviour of \hei-5876, while H$\alpha$ is shown in the bottom row. {The rms spectra show an inverted shape compared to the flux spectra, especially in the presence of P-Cygni profiles.} Statistical errors are smaller than the data points. Vertical dashed lines indicate the laboratory wavelength of the corresponding transition. Light yellow boxes highlight the regions affected by interstellar bands.}
\label{fig:line_comparison}
\end{figure*}
 
\section{Method}
\label{sec:method}
The estimator typically used to quantify the amplitude of variability in LMXBs is the rms. This is usually calculated by integrating the power spectral density of a time series, in specific units, while subtracting the contribution of Poisson noise \citep[see e.g.][]{belloni1990,Munoz-Darias2011,Munoz-Darias2014}. Since the consecutive spectra have significant readout gaps between them and may only cover flaring periods, these data are not ideal for Fourier domain analysis. Therefore, to quantify the variability of the spectrum, we computed the rms as the root of the excess variance for each wavelength element. This technique is often employed in time series analysis of Active Galactic Nuclei (AGN) and is defined as follows \citep[see also][]{vaughan2003}:
%{I REMOVED THE $\sigma^{2}_\mathrm{nxs}$ term in the equation because it is not expanded and hardly used}

\begin{equation}
rms_\mathrm{frac}=\sqrt{\cfrac{S^{2}-\overline{\sigma^{2}_\mathrm{err}}}{\overline{x}^{2}}}
\label{eq:ao}
\end{equation}

 where $S^2$ is the variance of a given time series $x(t)$,  $\overline{x}^{2}$ its squared average and $\overline{\sigma^{2}_\mathrm{err}}$ the average of the squared value of the errors.  Although the contribution of $\sigma^{2}_\mathrm{err}$ is negligible, the advantage of computing this quantity is that, unlike variance, it directly removes the contribution of Poisson noise, making it a more reliable proxy for intrinsic variability. Given the structure of the data (see Tab. 1 and Section 2), the described method evaluates the rms over Fourier frequencies of $\approx$2$\times$10$^{-2}$ Hz down to $\approx$1$\times$10$^{-4}$ Hz. We estimated the error on $rms_\mathrm{frac}$ following the equations described by \cite{vaughan2003}. Given the high S/N of the data, this can be done as follows:

\begin{equation}
err(rms_\mathrm{frac}) =  \sqrt{\frac{\overline{\sigma_{\rm{err}}^{2}}}{N}} \cdot\cfrac{1}{\bar{x}}
\label{eqn:xs_error}
\end{equation}

where N is the number is the number of spectra over which the rms is computed.

\section{Results}

\subsection{Overall behaviour}\label{s.overall}

The flux and rms spectra for each observing night are shown in Fig. \ref{fig:all}. A number of different properties can be distinguished:
\smallskip
\begin{enumerate}

\item[$\bullet$]  The mean level of variability changes significantly over the campaign (i.e. day 1 to 8). It is as high as $\sim 80$ per cent on day 2 and as low as $\sim$ 10 per cent on day 7 and 8, with the remaining observations in the range of $\sim$ 40 to 60 per cent. {In all the spectra, except those from day 3 and day 7, the rms spectrum exhibits a blue slope. Day 3 shows the most complex shape, characterised by ups and downs, while day 7 displays a red slope.}

\item[$\bullet$]The rms spectra show prominent dips corresponding to the presence of an emission line in the flux spectrum (see also Fig. \ref{fig:line_comparison}
), indicating that the emission components vary less than the continuum. This behaviour holds true for all lines except \heii\ at 4686 \AA\ (hereafter, \heii-4686) and the Bowen blend (i.e. C~$\textsc{iii}$ and N~$\textsc{iii}$ emission at $\sim$4640 \AA). These transitions, which are associated with the highest ionisation potentials, are observed to vary more than the adjacent continuum, resulting in emission peaks in the rms spectrum.
 
\item[$\bullet$] {The presence of a P-Cygni line profile in the flux spectrum leaves a distinct imprint on the rms spectrum: variability (i.e. rms) is enhanced in the blue part, while it decreases below the continuum level in the red part of the profile. Thus, when the flux and rms spectra are directly compared (e.g., top panels in Fig. \ref{fig:line_comparison}), the latter appears as an inverted version of the former, which we will refer to as an inverted profile in rms.} This is evident on day 2 and day 6, when the deepest blue-shifted absorptions (in flux) are detected in \hei-5876 (Figs. \ref{fig:all} and \ref{fig:line_comparison}). The inverted P-Cygni profile arises from the redshifted emission varying less than the continuum, consistent with the overall behaviour of the emission lines, and the blue-shifted absorption varying more than the adjacent continuum.

\item[$\bullet$] By examining {the \ha\ line in days 2 and 6}, we notice that the broad emission line detected in the flux spectrum appears as broad absorption in the rms spectrum (bottom panels in Fig. \ref{fig:line_comparison}
). However, the latter shows a strong asymmetry towards the blue, which can be attributed to the presence of an incipient emission component in this section of the line. As mentioned above, this component is clearly detected in \hei-5876. This suggests that even if the putative wind-related blue-shifted absorption is not strong enough to produce a dip in the flux spectrum (and thus a standard P-Cygni line profile), it can still leave an imprint in the rms spectrum, resulting in an asymmetry between the red and blue parts of the line.

\end{enumerate}

\begin{figure*}
\centering
\includegraphics[width=0.72\textwidth]{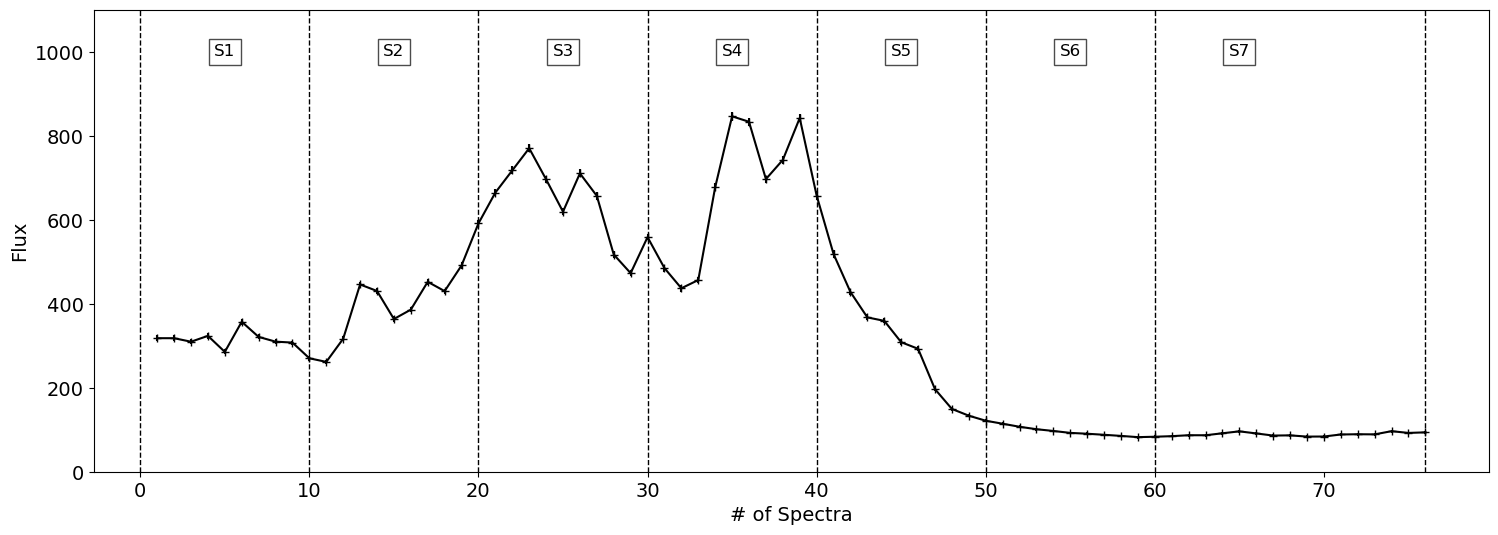}

\includegraphics[width=0.38\textwidth]{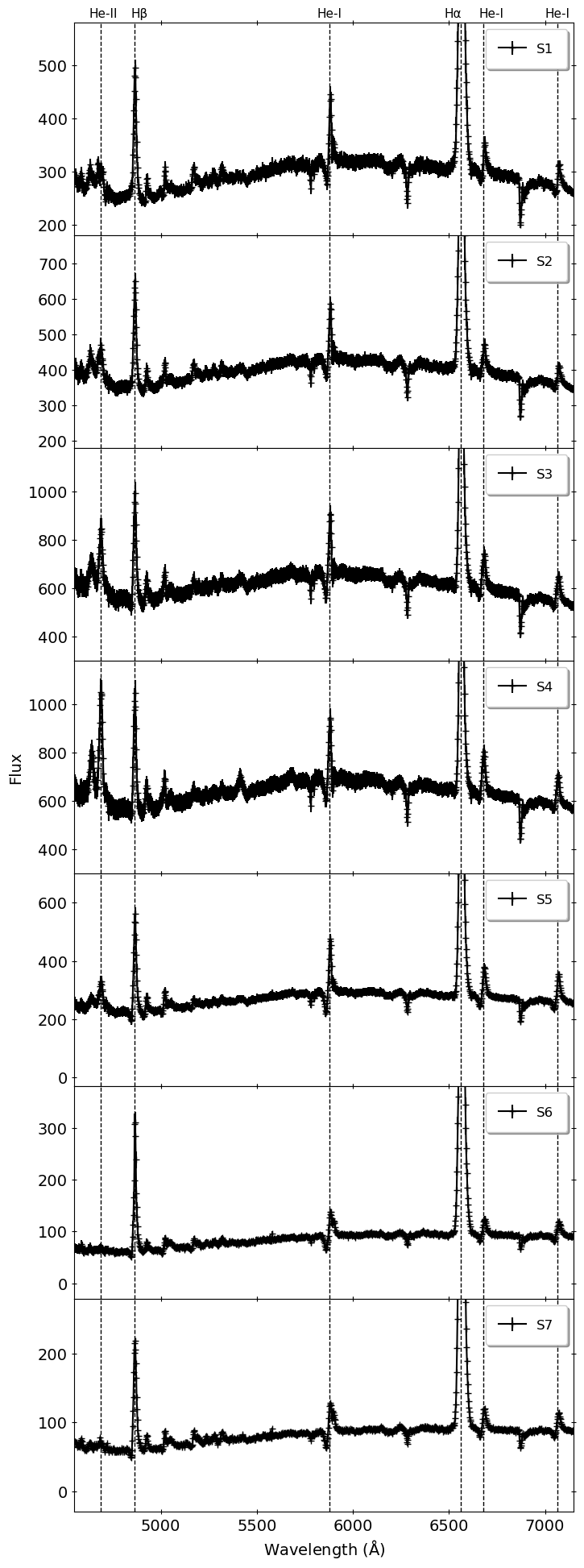}\hspace{0.5cm}
\includegraphics[width=0.38\textwidth]{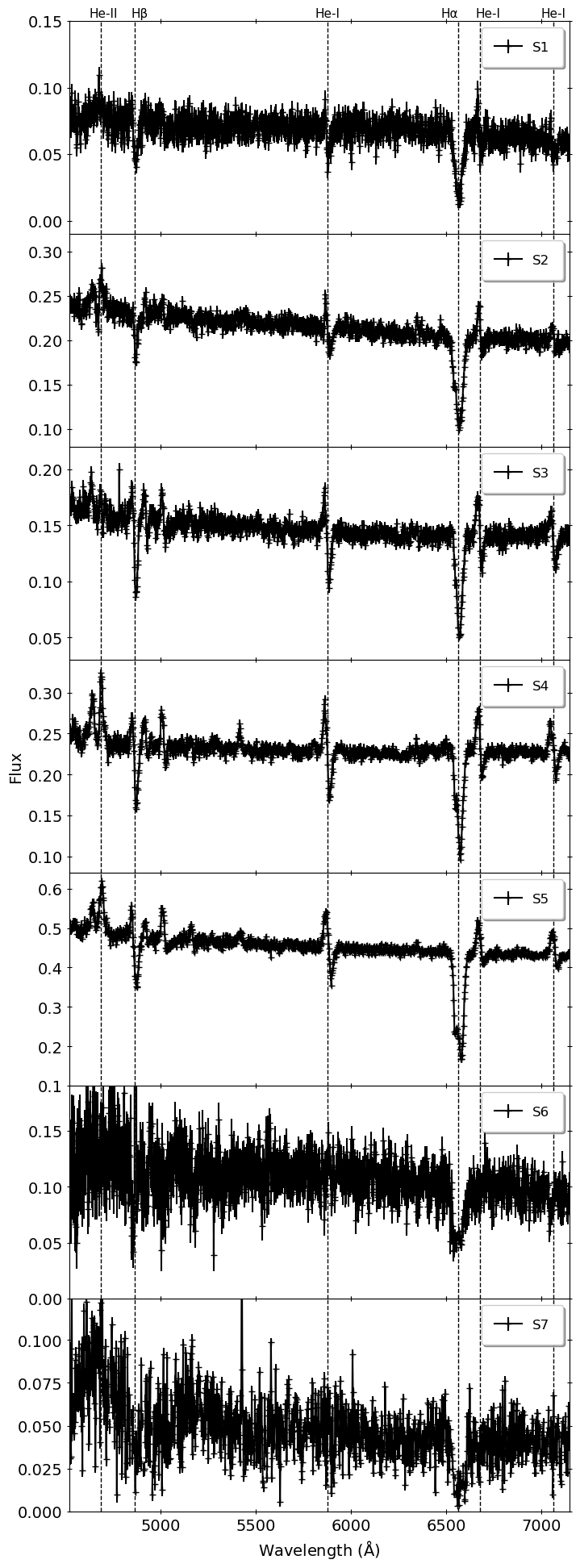}

\caption{Shorter-term evolution throughout day 2. The top panel shows the day 2 light curve and the intervals selected for calculating short-term rms spectra. The flux spectra for each segment are shown in the left column, with the corresponding rms spectra in the right panels. Dotted vertical lines mark the positions of the \hei\ lines (at $\sim$5876, 6678, and 7065 \AA), \heii\ 4686 \AA, \ha, and \hb. Interestingly, the \heii\ line shows significant flux variations between S3 and S5, coinciding with the strongest {inverted profiles} in the rms spectra.}
\label{fig:side}
\end{figure*}

\subsection{Shorter-term evolution of the rms spectra}
To gain further insight into the above-described behaviour, we also studied the short-term evolution of the rms spectrum. For this purpose, we selected day 2, as it is one of the longest segments and contains a glimpse of all key features, including a double-peaked optical flare. We divided the observation (76 spectra over 102 minutes) into six blocks of 10 consecutive spectra (S1 to S6), plus a final block containing the last 16 spectra at very low flux (S7), and computed the corresponding rms spectra. In this setup, S1 includes the first 10 spectra just before the rise of the flare, which is covered in S2, while S3 and S4 each sample one of the two peaks of the flare. S5 covers the abrupt decay from the peak to the flux minimum, and S6 and S7 cover the final part of the observation at the lowest flux levels. The results did not change when we slightly varied the number of spectra per segment. We note that in this short-term rms analysis, the low-frequency end of the band in which we are effectively computing the rms has shifted according to the shorter length of the individual blocks.

Fig. \ref{fig:side} shows the full day 2 light curve (top), along with the flux and rms spectra (left and right columns, respectively) corresponding to S1 through S7. The \hei-5876 P-Cygni line profile is visible in all the flux average spectra, {with the blue-shifted absorption becoming particularly deep (compared with the adjacent continuum) at low flux} (see also fig. 2 in \citealt{Munoz-Darias2016}). However, its rms counterpart (i.e., {the inverted profiles in rms}) changes in a more dramatic way throughout day 2. In particular, the wind-related feature is enhanced during segments associated with significant flux changes (S3 to S5), while it is very weak (S1 and S2) or absent (S6 and S7) when the flux remains consistently low. The positive and negative amplitude of the two components of the {inverted profile in rms} range from a few percent in S1 to nearly 10 percent in S5. A consistent behaviour is observed for the weaker helium lines (i.e., \hei-6678 and \hei-7065). Similarly, the absorption feature associated with \ha\ is observed to vary in a consistent manner, becoming stronger during the segments with larger flux variations. Notably, S4 and especially S5 show very deep absorptions, where the incipient blue emission component (see Sec. \ref{s.overall}) becomes evident.  In contrast, the \ha\ absorption is also visible in the S6 and S7 rms spectra, though the incipient blue emission is not apparent. 

The evolution of \heii\ in both sets of spectra also follows the changes in the day 2 light curve. While \heii\ is absent in both flux and rms spectra at the lowest flux levels (i.e., S1, S6, and S7), it becomes significantly strong in flux during the peaks of the flare (S3 and S4). In the rms spectra, on the other hand, it peaks during the segments associated with the largest variability in flux levels (S4 and S5) and is also present during the slow rise of the flare (S2). However, it is weak during S3, which primarily covers the first flare’s peak—a phase in which \heii\ remains strong but relatively constant.

Despite the points discussed above, it is important to note that the strength of wind-related features in the rms spectra (e.g., {the inverted profile in rms}) exhibits a complex dependence on both the average flux and the fractional rms (i.e., those associated with the continuum of the corresponding spectra). For instance, S2 and S4 exhibit similar rms levels, yet the amplitude of the {inverted profiles} differs significantly.

\section{Discussion}
The optical spectra displayed by the black hole transient V404 Cyg during its luminous 2015 outburst {arguably provide some of the best examples of optical wind signatures in LMXBs}. The quality of this unique dataset—the only one of its kind taken with $\sim$ minute sampling at high S/N—enabled tracking the observational properties of the optical wind on these short timescales \citep{Munoz-Darias2016}. Here, we take another step forward by studying, for the first time in this class of objects, the evolution of the optical rms spectra.

The study of the rms spectrum of LMXBs has already been attempted at other wavelengths. In particular, it is a relatively common method used to study different emission components through X-ray timing studies, a technique that has been successfully applied to several objects (see, e.g., \citealt{Gierlinski2005,Belloni2011}). At longer wavelengths, it was applied to ultraviolet spectra of the black hole transient XTE~J1118+480 \citep{hynes2003}, and more recently to mid-infrared observations of GRS~1915+105 \citep{gandhi2024}. Focusing on outflows, no wind-related signatures were observed in either the flux or rms spectrum in these studies. This is likely due to a combination of wavelength, accretion state, and, in the case of the rms, the small amount of variability characterising the observations.

Optical spectroscopic studies of discs and winds mostly rely on searching for and characterising the imprints they leave in spectral transitions, typically producing both absorption and emission components in the flux spectrum. The rms spectra allow us to quantify the amount of variability as a function of wavelength, and thus compare the variability levels of the different spectral components, such as spectral lines, both among themselves and against the continuum. Fig. \ref{fig:all} shows that the rms spectra are also rich in spectral features, though they are typically morphologically different from their counterparts in the flux. In particular, emission components in flux appear as absorption features in rms.
This means that even thought they are variable features, and they are seen to vary by more than $\sim$ 70 per cent (e.g. day 2 in Fig. \ref{fig:line_comparison}), they do it less than its adjacent continuum. However, this behaviour is not universal as transitions with higher ionisation potential, such as \heii-4686 and the Bowen blend are typically an exception to this rule as are also in emission in the rms spectrum. {The explanation for this behaviour is likely complex and multi-factorial. As an exploratory approach, we can consider photoionisation effects and the relative abundances of ions, which are key factors in producing the He and H recombination lines. For instance, in the range of ionisation parameters where different ions of H and He coexist ($\sim$0.5-50; see e.g., \citealt{kallman1982}), the He~\textsc{iii} ion (which recombines into \heii) is generally predicted to have a steeper relative abundance curve than other relevant H and He ions. Therefore, small changes in the incident radiation (i.e., variability) would lead to significant changes in the relative ion abundance and, consequently, in the strength of the \heii\ lines. Following this argument, a relatively flat abundance curve would dampen the impact of the incident variability, resulting in smaller changes in the strength of the line, which would explain the behaviour of the \hei\ and Balmer lines. However, we note that the above is strongly dependent on several factors, including (but not limited to) the densities and luminosities involved.}

The strength, and even the mere presence, of blue-shifted absorptions associated with optical P-Cygni line profiles in LMXBs has generally been observed to be highly variable in the few datasets where sufficient individual spectra have been collected (see, e.g., \citealt{Munoz-Darias2020, cuneo2020, Charles2019, Jimenez-Ibarra2019b} for other sources).  In the case of V404 Cyg (i.e., this dataset), the presence or absence of this feature (mainly in the Balmer and \hei\ transitions) correlates well with the ionisation state, measured as the H$\beta$ to \heii\ flux ratio \citep{Munoz-Darias2016}.  Thus, P-Cygni line profiles become much weaker as ionisation increases (i.e., as \heii\ becomes stronger). It is therefore not surprising that these lines behave similarly to \heii\ in the rms spectra, varying more than the adjacent continuum. 

We should bear in mind that producing the blue-shifted absorption part of the P-Cygni profile requires significant wind-related scattering along the observer’s line of sight. Often, this scattering is only sufficient to make the emission component asymmetric towards the red (i.e., flux is reduced in the blue part, but no absorption component forms). This makes the feature particularly sensitive to the ionisation field and, consequently, to the relative abundances of ions involved in the transition. Specifically, strong P-Cygni line features require the lower level of the transition to act as the effective ground state, from which an electron continuously excites and recombines. This makes these features highly sensitive to the physical conditions of the gas. {In summary, our results can be understood as a direct consequence of the general picture originally proposed from this dataset \citep{Munoz-Darias2016}, which suggests the continuous presence of an outflow, with its visibility varying according to the physical conditions of the ejecta. This interpretation is further supported by independent analyses, using different techniques, of ultraviolet spectral variability from the neutron star LMXB Swift J1858.6-0814 \citep{castro-segura_2022}.}

\begin{figure*}
    \centering
\includegraphics[width=0.45\textwidth]{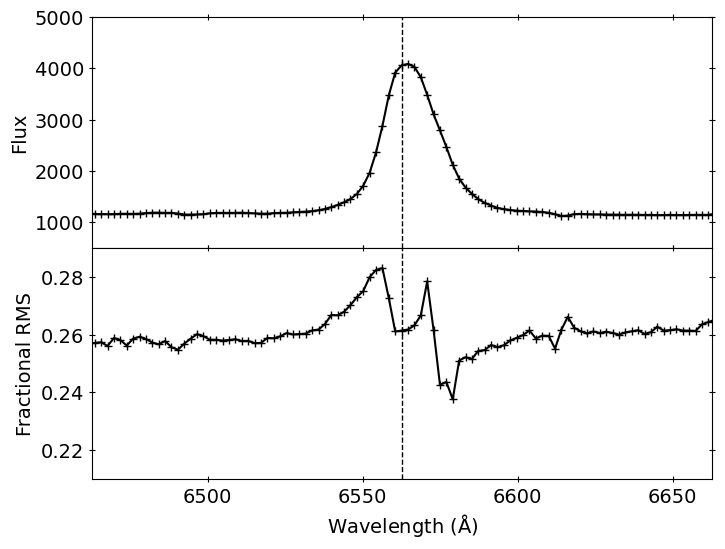}
\includegraphics[width=0.45\textwidth]{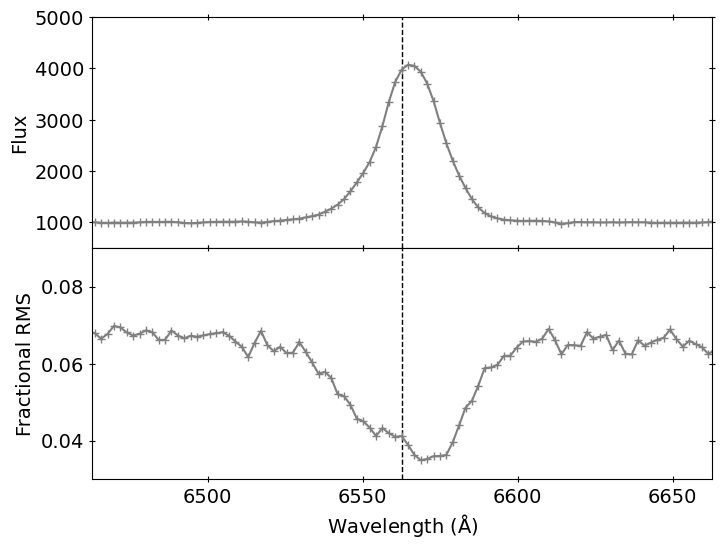}
 
\caption{Detailed rms and flux spectra of \ha\ for day 7 (left) and day 8 (right). While the flux absorption in the blue part of the line is mild, the asymmetry in the rms spectrum is evident.}

\label{fig:line_comparison_day7-8}
\end{figure*}

\subsection{The combined use of flux and rms spectra}

Our results show that rms spectra can be a useful tool for studying accretion disc winds. In particular, when combined with high S/N flux spectra (including those obtained by averaging several individual spectra), this approach increases the sensitivity of the analysis by (i) enhancing the confidence of detection when both P-Cygni (in flux) and {inverse profiles (in rms)} are observed, and (ii) revealing weak features when asymmetries are present in the emission (flux) and absorption (rms) line profiles.

To illustrate (i), we note that it is often the case that a weak blue-shifted absorption is seen in a particular line. Its origin may be related to the presence of wind-type ejecta, but it can also be attributed to interstellar or telluric absorption (e.g., the H$\alpha$ line can be affected by weak sky and interstellar bands on both sides), or artefacts introduced during data reduction and normalisation. This can be addressed by searching for a similar feature in a different transition or by making further considerations about the line profile, including applying more complex machine learning techniques (e.g., \citealt{matasanchez2023}). However, a detailed inspection of the rms spectra reveals that features unrelated to the source do not leave an imprint in the rms spectra, nor do they mimic the behaviour of {an inverted profile in rms}. An example of this can be seen in Fig. \ref{fig:line_comparison} (top panel), where the interstellar bands present in the blue continuum and the Na doublet affecting the red wing of the \hei\ line (marked with yellow bands) do not impact the rms spectrum. 

Likewise, to illustrate (ii), we can focus on day 7 and day 8, when P-Cygni line profiles are not detected in the flux spectra (see Fig. \ref{fig:line_comparison_day7-8}). 
However, the corresponding H$\alpha$ line profile in the rms spectra reveals an asymmetry in the absorption component (see bottom panels in Fig. \ref{fig:line_comparison_day7-8}). This component is strong on day 7, where the blue part of the line appears in emission, {resembling an inverted profile in rms}. We also note that in addition to the asymmetry, an emission component embedded within the broad underlying absorption is consistently present in H$\alpha$, particularly on day 7, when the highest flux of our campaign was recorded (see also bottom panels in Fig. \ref{fig:line_comparison}). This could be due to the contribution of the accretion disc to the line flux. Thus, during the brightest day (day 7), this component would be enhanced. We note that while previous studies on this data set \citep{Munoz-Darias2016, mata-sanchez2018, Casares2019, Munoz-Darias2022} suggest that optical lines are mainly formed in the wind, the presence of a disc component partially contributing to the line flux is not unexpected.

The above-described synergy between flux and rms spectra becomes evident when examining the evolution of the short-term behaviour in flux and rms during day 2 (see Fig. \ref{fig:side}). The strongest wind signatures in the rms spectra are observed in S2 through S5, which correspond to segments with variability in the day 2 lightcurve. In these data intervals, {inverted profiles in rms} are associated with the \hei\ lines (at 5876, 6678 and 7065 \AA) as well as H$\beta$. For H$\alpha$, the absorption feature becomes markedly asymmetric due to the presence of incipient blue-shifted emission embedded within the broad absorption component. However, when looking at the flux spectra, P-Cygni line profiles are either not seen or are much weaker during these data intervals, with the deepest blue-shifted absorptions detected in S6 and S7, when the flux is low (see also fig. 2 in \citealt{Munoz-Darias2016}). Since the lightcurve is stable during these segments (i.e. S6 and S7), the rms spectra are not particularly revealing, despite strong wind signatures being observed in the flux counterparts. This highlights that a deeper search for winds can be achieved by combining rms and flux spectra—a strategy that should be considered in future observing campaigns, particularly for relatively bright sources where a large number of spectra can be obtained within a single observing window.

\subsection{Rms spectra and winds beyond X-ray binaries}

Several studies of different accreting systems have already indicated that rms spectra can be a powerful tool for detecting and characterising outflows. This method was first developed to characterise the long-term, energy-dependent variability of active galactic nuclei (AGN) using non-homogeneous sampling \citep{edelson,vaughan2003}. Further application of this method to high-quality XMM-Newton data led to the detection of excess variability around the iron line in candidates thought to harbour ultra-fast outflows \citep[e.g.,][]{parker2017,igoz}. Detailed modelling of this behaviour has shown that this phenomenon might be linked to variations in the ionisation of the wind \citep{pinto2018}. Our results are consistent with a similar physical mechanism impacting the optical lines arising in the wind of the black hole transient V404 Cyg. Given the robustness of the wind detection in this object, our findings reinforce the value of rms spectra-based techniques for wind searches.

{From a theoretical perspective, P-Cygni line profiles have been explored for many years in massive stars, as well as in other objects such as accreting white dwarfs (see e.g., \citealt{castor1979,Shlosman1993,Lamers_Cassinelli_1999}), which also allow the study of different wind geometries.} Variations in the P-Cygni profile can arise due to several factors, including changes in optical depth, fluctuations in wind mass loss, or perturbations in the wind structure. Indeed, several observational studies have reported the presence of variable ultraviolet and optical P-Cygni line profiles in a number of accreting white dwarfs \citep[see, e.g.,][]{Drew1997,patterson2001,cuneo2023}. However, fractional rms spectra have not yet been used as a diagnostic tool to probe the physical parameters of these outflows. In light of our results, and given the relatively large sample of optically bright accreting white dwarfs available, rms spectroscopy appears to be a promising avenue for gaining new insights into the observational properties of accretion disc winds in these objects \citep[see also][]{georganti2022}. This approach should not necessarily be limited to improving wind detectability but can also help explore the physical parameters of these outflows when combined with multiwavelength spectral information and modeling.

\section{Conclusions}

We presented, for the first time, optical rms spectra of a black hole transient in outburst. In particular, we analysed data from the remarkable 2015 outburst of V404 Cyg, where clear signatures of wind-type outflows were identified in several emission lines. Using this unique dataset, {we show that winds leave an imprint in the rms spectra in correspondence to the standard P-Cygni line profiles observed in the flux spectra. This imprint takes the form of an inverted profile in the rms spectrum, characterised by enhanced variability in the blue-shifted region accompanied by a red absorption component.} Our analysis reveals that the intensity of these features is closely linked to overall flux variation, suggesting that rapid changes in the ionisation state of the gas play a significant role in the formation of the {inverted profiles} in the rms spectra. Similar behaviour has been observed in AGN, where this technique has been applied to study ultra-fast outflows. Given the robustness of the wind detection in our study of V404 Cyg, our results reinforce the value of rms spectra-based techniques for wind searches. We emphasise that, given the ubiquitous presence of outflows across accreting systems, variability studies using rms spectra represent a powerful and versatile approach to exploring wind phenomena across a wide range of astrophysical objects.

\begin{acknowledgements}
We thank the referee for the useful suggestions that helped to improve the quality of the paper. FMV thanks C. Knigge for the useful discussion regarding the intepretaion of the rms  spectrum and acknowledges support from the Science and Technology Facilities Council grant ST/V001000/1. T.M.-D. is thankful to Z.~Igo for useful discussions on the use of rms spectra in AGN during the second SHIVA meeting (Umbria, Italy, 2024). T.M.-D. also acknowledges support from the Spanish \textit{Agencia Estatal de Investigación} via PID2021-124879NB-I00.
\end{acknowledgements}
%%%%%%%%%%%%%%%%%%%%%%%%%%%%%%%%%%%%%%%%%%%%%%%%%%
\bibliographystyle{aa} 
\bibliography{biblio}

%\listofobjects 

\end{document}